\documentclass[aps,showpacs,preprintnumbers,amsmath,amssymb,superscriptaddress,floatfix,a4paper,nofootinbib,11pt,floatfix,nofootinbib,onecolumn]{revtex4}
\usepackage{graphicx}
\usepackage{bm}
\usepackage{rotating}
\usepackage{array}
\usepackage{xcolor}
\usepackage{amsmath,amssymb}
\usepackage{mathrsfs}
\usepackage{graphicx}
\usepackage{color}
\usepackage{subfigure}
\usepackage{fancyhdr}
\usepackage{multirow}
\usepackage{float}
\usepackage{epsfig}
\usepackage{amsfonts}
\usepackage{bm}

\newcommand{\ba}{\begin{eqnarray}}
\newcommand{\ea}{\end{eqnarray}}

\begin{document}

\title{Composite Nambu--Jona-Lasinio inflation near infrared fixed point of the Ho\v{r}ava-Lifshitz theory}

\author{Daris Samart} 
\email{dsamart82@gmail.com}
\affiliation{High Energy Physics Theory Group, Department of Physics, 
\\Faculty of Science, Chulalongkorn University, Phyathai Rd., Bangkok 10330, Thailand}

\author{Phongpichit Channuie} 
\email{channuie@gmail.com}
\affiliation{College of Graduate Studies, Walailak University, Thasala, Nakhon Si Thammarat, 80160, Thailand}
\affiliation{School of Science, Walailak University, Thasala, \\Nakhon Si Thammarat, 80160, Thailand}
\affiliation{Thailand Center of Excellence in Physics, Ministry of Education, Bangkok 10400, Thailand}

\date{\today}

\begin{abstract}

In this work, we first propose a cosmological scenario inherently based on the effective Nambu--Jona-Lasio (NJL) model near the infrared (IR) limit of the Ho\v{r}ava-Lifshitz (HL) theory. Having used the one-loop correction, we employ the NJL framework in the ultraviolet (UV) limit of the HL theory, with critical exponent $z=3$, to demonstrate in the IR limit that the symmetry of the potential at the critical coupling value is broken at $z=1$. We also derive the effective Higgs-like potential in the low energy regimes. Here the symmetry of the effective potential will be broken near $z=1$ at a certain value of the critical coupling. In our scenario, the scalar channel of the NJL model plays the role of a composite inflaton. We find that the Lorentz invariance (LI) appears at the IR regime of the HL theory and employ the standard inflationary (slow-roll) paradigm as a probe of physics at very high energy. We compute inflationary parameters and compare them with recent Planck 2015 data. Interestingly, we discover that the predictions of the model are in perfect agreement with the Planck analysis. Our salient feature is that we used inflation to quantify the IR and UV fixed points of underlying theory. We discover that the IR fixed point of HL gravity is compatible with the grand unified energy scale; whilst the UV fixed point is the Planck energy scale.

\end{abstract}


\maketitle


\section{Introduction}
Quantum theory of gravity is one of the unsolved problems in theoretical physics and there is enormous uncertainty about its nature. It is strongly believed that it might deserve as the theory of everything and would be expected to describe physics of the singularity inside black holes as well as the origin of the universe or the big bang. In the past decade, Ho\v{r}ava proposed an alternative description of the quantum theory of gravity by introducing anisotropic scaling of time and space with the critical exponent parameter, $z$, to achieve the renormalizability of quantum gravity at the ultraviolet (UV) limit \cite{Horava:2009uw}. 

Although this method leads to the violation of the Lorentz invariance (LI), i.e., a breakdown of standard space-time structure at small distances, there is no experimental evidence, or constraint, of the Lorentz symmetry at extremely high energies \cite{Horava:2009uw}. However, the LI in the Ho\v{r}ava gravity is inherently restored at low energy regimes of the theory. It is worth noting that it was proposed by Horava and used Lifshitz invariance to quantize gravity without invoking any exotic particles or extra dimensions. There are many fruitful applications and consequences of the HL theory in a context of astrophysics (black hole), particle physics and cosmology (see Refs.\cite{Farias:2011aa,Cognola:2016gjy,Saridakis:2009bv,Bogdanos:2009uj,Dutta:2010jh,Dutta:2009jn} and Ref.\cite{Wang:2017brl} for a recent review).

At a very high energy scale, the mechanism responsible for an early rapid expansion of our Universe is well accepted nowadays since its predictions are in excellent agreement with the present observations \cite{Ade:2015lrj}. This period of such expansion is known as \lq\lq cosmic inflation\rq\rq \cite{Alex,KSa,KSa1,DKa,GUT}. However, little is known about the mechanism underlying the inflationary physics. The simplest models of inflation make use of elementary
scalar fields. It was pointed out in Ref.\cite{Bezrukov:2007ep} that the inflationary mechanism in the early Universe can be connected to the standard model (SM). This was achieved just by identifying the scalar Higgs boson with the inflaton. 

The salient feature of the Higgs-inflation paradigm is the requirement of the non-minimal coupling of the Higgs doublet field $(H)$ to gravity, i.e., $\xi H^{\dagger}HR$ with $\xi$ a new coupling constant.  A large non-minimal coupling is required in the Higgs-inflation scenario, $\xi \sim 10^{4}$. However, the elementary scalar field in field theories is defective in the sense that quantum corrections generate unprotected quadratic divergences
which must be fine-tuned away. One of the solutions involves a composite field of some strongly coupled theories (see \cite{Hill:2002ap} for a review) which feature only fermionic matter, and therefore the theory is stable with respect to quantum corrections. 

In addition, in the effective Lagrangian description for light mesons, the Nambu-Jona-Lasinio (NJL) model \cite{Nambu:1961tp,Nambu:1961fr} is another time-honored example. Recently, the authors of \cite{Inagaki:2015eza,Channuie:2016iyy} engaged the NJL model with the slow-roll inflationary scenario to show that the successful inflation can be nicely achieved. See also Refs.\cite{Evans:2010tf,Channuie:2011rq,Bezrukov:2011mv,Channuie:2012bv} for other composite field scenarios. In the composite field scenario, a non-minimal coupling of the composite scalar field $\Phi$ to gravity can be naturally induced \cite{Hill:1991jc}. Moreover, as suggested in Ref.\cite{Hill:1991jc}, if $\xi$ is a running constant, its value at the NJL cutoff scale $\Lambda$, $\xi(\Lambda)$ might be large and, then evolve toward $\xi = 1/6$ at lower energies. 

Moreover, the NJL model is very compelling when one wants to examine the phenomena of spontaneous symmetry breaking. This is frequently used to explain the spontaneous generations of the particle masses by applying the machanism of symmetry breaking, e.g., chiral symmetry breaking of hadron masses in QCD, and electroweak symmetry breaking of gauge bosons in standard model of particle physics. The NJL model has also been studied in the Lifshitz-like theory at $z=3$ \cite{Dhar:2009dx}, and the results showed that the model is asymptotically free, and mass is generated dynamically. 

In addition, the composite Higgs field can emerge in this scenario. Here, the constituent fermion mass of the condensation can be generated for arbitrary coupling constant of the theory. Interestingly, this phenomenon has the same profile as that of the NJL model in holographic Sakai-Sugimoto (AdS/QCD) scenario \cite{Antonyan:2006vw}. But in the latter, we will show how the symmetry breaking of the model at some certain value of the coupling constant in the IR limit of the Lifshitz-like theory can happen like a standard relativistic NJL model. Moreover, the Einstein static universe (ESU) and the emergent universe scenario in the framework of Hoˇrava-Lifshitz-like $F(R)$ gravity can be found in Ref.\cite{Khodadi:2015fav}.

In this work, we will investigate composite NJL inflation near the IR limit of the HL theory in which the effective potential can be obtained using the one-loop expansion scheme. Note that the HL gravity converts to GR in this regime, i.e., $z\rightarrow 1$. One therefore considers the inflationary phase of the universe as the IR limit of the HL theory. More importantly, we first demonstrate the interval of UV and IR regimes of the HL theory via the inflationary constraints.
\section{Composite NJL scenario and its effective potential}
We start with the action of the NJL model in the HL-like theory at $z=3$. It reads \cite{Dhar:2009dx},
\begin{eqnarray}
S &=& \int d^4x\, \Big[
i\,\bar \psi_{i}
\left( \gamma^0\partial_t + \vec \partial \cdot {\vec \gamma} \ \partial^2\right) \psi_{i}
+ g^2 \ \bar\psi_{L,i}\,\psi_{R,i} \,\bar\psi_{R,j}\, \psi_{L,j} \Big],
\label{four-fermi}
\end{eqnarray}
where $d^4x \equiv dtd^3\vec x$, $\partial_t \equiv \partial/\partial t$, $\gamma^0$ and $\vec \gamma$ are standard Dirac matrices in time and spatial components, respectively. The left- and right-handed fermions are defined by $\psi_{L/R}= P_{L/R}\psi = (1\pm \gamma_5)\psi/2$ and $g$ is the coupling constant of the NJL interaction term. The model consists of the color $SU(N_c)$ and the fermions $\psi_{i}(t,\vec x)$ transform vectorially as follows:
\begin{eqnarray}
\psi_{i} \to U_{ij}\psi_{j}
\,,\qquad i,j =1,\cdots,N_c\,.
\end{eqnarray}
Note that additional symmetries of this theory are discussed in detail in Ref.\cite{Dhar:2009dx,Dhar:2009am}. For simplicity, we focus only on the global $U(N_c)$ symmetry which is associated with the large-$N_c$ approximation. At the UV limit of the HL theory, the space and time can be transformed by using an anisotropic scaling of a critical exponent $z$ as,
\begin{eqnarray}
\vec x \rightarrow b \vec x\,, \qquad\quad t \rightarrow b^3 t\,.
\label{scale}
\end{eqnarray}
In the case of the HL-like theory with $z=3$, we then obtain scaling dimensions of time and space components in terms of mass dimension as $[L]= -1$, $[T]=-3$, and theses give $[\psi]= 3/2$ and $[g]=0$. Then the action in Eq.(\ref{four-fermi}) is of dimension 6. In addition, the $g$ coupling is dimensionless which is contrary to the traditional NJL theory at Lorentz invariant regime, i.e., $[g]_{z=1} = -2$. By using a standard method of bosonization, we can rewrite the NJL interaction in terms of auxiliary ``composite" fields, $\Phi$ as
\begin{eqnarray}
S &=& \int d^4x \ \Big[
i\,\bar \psi_{i}
\left( \gamma^0\partial_t + \vec \partial \cdot {\vec \gamma} \ \partial^2\right) \psi_{i}
- \frac{1}{g^2} \Phi^\dagger\Phi 
+ \Phi^\dagger \bar\psi_{R,i}\psi_{L,i} + \Phi\, \bar\psi_{L,i}\psi_{R,i} \Big]\,.
\label{NJL-com}
\end{eqnarray}
Integrating out the fermion fields in Eq.(\ref{NJL-com}) and using the standard method of a one-loop expansion, we obtain the effective potential as \cite{Dhar:2009dx}
\begin{eqnarray}
V_{\rm eff}(\Phi)=  \frac{|\Phi|^2}{g^2}\left(1- \frac{\lambda_N}{12\pi^2}
\left[1 + \ln\left(\frac{\Lambda^6}{|\Phi|^2} \right) \right] \right),
\label{eff-pot}
\end{eqnarray}
where $|\Phi|^2 \equiv \Phi^\dagger\Phi$, $\lambda_N \equiv g^2 N_c$ and $\Lambda$ are a 't Hooft coupling and a 3-momentum UV cut-off, respectively. It is simple to obtain in Eq.(\ref{eff-pot}) that $[ \Phi ] = 3$\,,  $[ \lambda_N ] = 0$ and $[\Lambda] = 1$. In this theory, the relevant coupling is $\lambda_N$ which has as asymptotic freedom behaviour \cite{Dhar:2009dx}. Moreover, by using standard renormalization group analysis, the coupling behaviour can be figured out as well as the fate of chiral symmetries, see Ref.\cite{Dhar:2009am}. As mentioned earlier, it is worth noting that the spontaneous symmetry of the effective potential in Eq.(\ref{eff-pot}) always breaks at arbitrary weak coupling values, $g$ \cite{Dhar:2009dx}. This differs from the standard NJL model, and we will show in the latter how to break the symmetry with some critical value of the coupling $g$\,. 

Before proceeding further, it is 
worth demonstrating how the LI of the HL theory is restored. Here we will briefly repeat the mechanism of emergence of LI near IR fixed point. Considering the scalar field $\Phi$, one can parametrize the fluctuations in terms of a radial field $(\varphi)$ and a phase $(\pi)$ pseudoscalar field: 
\begin{eqnarray}
\Phi(x) = \rho(x) e^{i g\,\pi(x)}, \qquad\quad  \rho(x) = m^3+ g\, \varphi(x)\,.
\label{def-pi}
\end{eqnarray}
Moreover, introducing the fermion-mass condensation, $m$, via $ m^6 \equiv <0||\Phi|^2|0>$, it has been shown in Ref.\cite{Dhar:2009dx} that the LI of the HL theory is recovered. Here we will briefly repeat the mechanism of emergence of the LI. Firstly, we consider the action in Eq.(\ref{NJL-com})  by introducing the effective-mass squared $M^2$ into the kinetic term, $\bar\psi\,i\,(\vec\gamma\cdot\vec\partial\,M^2)\psi$\,. Then the action can be re-written to obtain
\begin{eqnarray}
S &=& \int d^4x \Big[
\bar\psi_i
\left(i \gamma^0\partial_t + i (\vec \partial \cdot \vec \gamma\,)\big( \vec\partial^2 + M^2 \big)\right) \psi_{i}
- \frac{1}{g^2} \Phi^\dagger \Phi +  \bar\psi_i\left( \Phi\,P_L + \Phi^\dagger\,P_R \right) \psi_i
\Big]\,,
\label{NJL-LI}
\end{eqnarray}
where $P_{L/R} = \frac12\,(1\pm \gamma_5)$. Substituting the $\Phi$ field (\ref{def-pi}) to the action (\ref{NJL-LI}), the on-shell mass equation in the kinetic terms of the fermion fields in Eq. (\ref{NJL-LI}) is given by
\begin{eqnarray}
k_0^2 - k^2\left(M^2 + k^2\right)^{2} - m^6 = 0\,,
\end{eqnarray}
where $k_0$ and $k$ are a momentum of the field $\Phi$ in time and spatial components, respectively. At the low momentum limit $(k \ll M)$, and re-scaling the energy $(k_0\rightarrow k_0/M^2)$\,, one recovers the LI (relativistic energy-momentum relation) as,
\begin{eqnarray}
k_0^2 = k^2 + m_*^2\,,\quad m_*\equiv m^3/M^2\,. 
\end{eqnarray} 
Here we will identify the parameter $M$ as the effective cut-off near the IR limit, and the LI appears below the momentum scale $M$ (see section IV of Ref. \cite{Dhar:2009dx} for detail discussion). The parameter $M$ is regarded as the RG-invariant mass scale which is identified as a cut-off dependent coupling $M = M(\Lambda)$ \cite{Dhar:2009dx}. We therefore introduce the cut-off dependent parameter $M(\Lambda)$ as a linear function of the UV cut-off $\Lambda$, {\it viz.},
\begin{eqnarray}
M(\Lambda) = \Lambda/\alpha  \,,
\label{cut-off-ansatz}
\end{eqnarray}
where $\alpha$ is a new dimensionless parameter which displays how far the UV and IR cut-off of the theory are separated. We will see in the next section that the ansatz in Eq. (\ref{cut-off-ansatz}) will be very useful for identifying the UV and IR limits of the HL theory when we consider inflationary parameters. Now we demonstrate the symmetry breaking phase of the $\Phi$ field that occurs in the critical value $g_c$ of the NJL coupling. To do this, we take the re-scalings of the field and the effective potential in Eq. (\ref{eff-pot}) such that
\begin{eqnarray}
\Phi \to M^2\Phi\quad\quad{\rm and}\quad\quad V_{\rm eff}\to V_{\rm eff}/M^{2}.
\end{eqnarray}
Assuming at the unbroken symmetric phase, in addition, the fermion condensate mass is vanished ($m=0$) then we can use the approximation $\Phi(x) \approx g\, \varphi(x)\,e^{i g\,\pi(x)}$. Using the mentioned re-scaling variables and $\Lambda = \alpha\,M$ ansatz, the effective potential near the IR fixed point in the unbroken phase is given by,  
\begin{eqnarray}
V_{\rm eff}(\varphi)=  M^2\varphi^2\left(1- \frac{\lambda_N}{12\pi^2}
\left[1 - \ln\left(\frac{g^2\varphi^2}{\alpha^6 M^2} \right)\, \right] \right).
\end{eqnarray}
We can further simplify the above potential by applying standard Taylor expansion such that $\ln (1+x) = x + \mathcal{O}\big( x^2\big)$. Replacing $1+x$ with $x$, therefore, the Higgs-like effective potential near the IR limit takes the form,
\begin{eqnarray}
V_{\rm eff}^{IR}(\varphi) &=& M^2\varphi^2\left(1- \frac{\lambda_N}{12\pi^2}
\left[1 - \ln\left(\frac{g^2\varphi^2}{\alpha^6 M^2} \right)\, \right] \right) \nonumber\\ &=&M^2\varphi^2\left(1- \frac{\lambda_N}{12\pi^2}
\left[1 - \left(\frac{g^2\varphi^2}{\alpha^6 M^2} -1  + \mathcal{O}\Big(\frac{g^2\varphi^2}{\alpha^6 M^2} - 1\Big)^{2}\right)\, \right] \right) \nonumber\\ &\approx& -\widetilde m^2\varphi^2 + \lambda\,\varphi^4 \,,
\label{Higgs-like-pot}
\end{eqnarray}
where the above expansion is still valid when an argument  $\frac{g^2\varphi^2}{\alpha^6 M^2}$ is close to one and we have kept only the first-order term for further examination. Here $\widetilde m$ and $\lambda$ parameters are defined by
\begin{eqnarray}
\widetilde m^2 = M^2 \left(\frac{g^2 N_c - 6\pi^2}{6\pi^2} \right), \qquad 
\lambda = \frac{g^4N_c}{12\pi^2\alpha^6} \,.
\label{couplings-higgs-pot}
\end{eqnarray}
More importantly, it has been pointed out in Ref. \cite{Dhar:2009dx} that the scalar field $\Phi$ might play the role of the composite Higgs fields in the low energy limit of the HL theory, i.e. $z\rightarrow 1$. In addition, it is easily shown that the coupling, $g$, is no longer a weak coupling. Moreover, one finds that the symmetry will be broken when,
\begin{eqnarray}
g^2  > \frac{6\pi^2}{N_c} \equiv g_c^2  \,.
\label{critical-g}
\end{eqnarray}
In the next section, we propose inflationary scenario by using the Higgs-like potential in Eq.(\ref{Higgs-like-pot}) from the HL theory. Note here that the study of the critical value of the coupling $g$ when the LI emerges is worth for the future investigations.

\section{Inflationary implications of the HL theory near IR fixed point}
This section aims to study inflation and employs the effective potential in Eq.(\ref{Higgs-like-pot}). We will see that a form of the Higgs-like potential is suitable for inflation. Additionally, the potential in Eq.(\ref{Higgs-like-pot}) is obtained near the IR limit where the LI is restored. Notice from Refs.\cite{Kluson:2011ff,Kluson:2011qv,Moon:2010wq} that when performing a conformal transformation of the gravitational action at $z=D=3$, there is a conformal factor left in front of the Ricci scalar, e.g., $f(\Omega){\tilde R}$, of the resulting action, and hence we cannot end up with the canonical Einstein-Hilbert gravitational action, a.k.a. the Einstein frame. We refer the readers to Refs.\cite{Kluson:2011qv,Kluson:2011ff,Moon:2010wq} for detailed derivations and its implications in $f(R)$ and scalar-tensor theories. In addition, Refs.\cite{Kiritsis:2009sh,Calcagni:2009ar} have demonstrated that the power spectrum is not scale invariant with the standard inflaton (scalar) field in the UV fixed point. On one hand, this means that anisotropic scalings of the HL theory at the critical exponent $z=D=3$ might not be correct in the inflationary phase or the inflation of the universe prefers the LI to generate a correct scale invariant of the power spectrum. On the other hand, we need to modify the scalar field sector. From these points, in this work, it is reasonable to consider the inflationary phase as the IR limit of the HL theory where the LI is restored. Furthermore, a so-called matter bounce scenario in the IR limit of the HL theory was studied in Ref.\cite{Brandenberger:2009yt} as an alternative to inflation for producing a scale-invariant spectrum of cosmological perturbations. We therefore study the effective potential of the composite scalar field of the HL theory near the IR fixed point in Eq.(\ref{Higgs-like-pot}) as the inflaton potential. In addition, we will consider the gravitational action near the IR fixed point of the HL theory, i.e. $z\rightarrow 1$ limit, where isotropic scalings and LI are recovered and the inflationary phase is an effective low energy scale for the UV fixed point (the Planck scale). 

In so doing, we adopt the same manner for the conformal transformation of the non-minimal coupling of the HL gravity from Ref.\cite{Kluson:2011ff} with $z\rightarrow 1$ limit. This helps to transform the action from the Jordan to the Einstein frame similar to the standard cosmology. By using the both of the effective potential and the gravitation action in the IR limit of the HL theory, therefore, we can safely use the conformal transformation in the GR case as proposed in Ref.\cite{Bezrukov:2011mv}. Here we introduce the non-minimal coupling term, $\xi\,\Phi^{2/d}\,\mathcal{R}$, to the action of HL theory in the Jordan frame with the inflaton denoted as $\Phi$. This corresponding counterterm was introduced in Ref.\cite{LopezNacir:2011mt}. This field
has mass dimension $d$ and couples to gravity via the non-minimal coupling term near the IR fixed point in the HL framework. By using one-loop renormalization in curved backgrounds, more importantly, it was mentioned in Ref.\cite{LopezNacir:2011mt} that a large non-minimal coupling between a scalar field and curvature (graviton) can be in principle generated if the scale $M\ll \Lambda$. Here the scale $M$ could be interpreted as the IR regime (i.e. $z\rightarrow 1$); while the Lorentz violation scale $\Lambda$ could be interpreted as the a UV regime (i.e. $z = 3$). Therefore it is reasonable to consider the case with $z=1$ in order to study inflation in the HL theory. Firstly, we recall the effective gravitational action of the HL theory in ADM form at the IR limit written as \cite{Blas:2009qj}
\begin{eqnarray}
S = \frac{1}{16\pi\,G}\int dt\,d^3x\,N\,\sqrt{h}\,\Big[ (1 - \alpha)\,K_{ij}\,K^{ij} - (1 + \lambda_{HL})\,K^2 + \, {}^{(3)}R + \beta\,a_i\,a^i\Big]\,,
\label{ADM-non-project-HL}
\end{eqnarray}
where ${K}_{ij}$, ${K}$ and $\, {}^{(3)}{R}$ are the extrinsic curvature, its trace and Ricci-like scalar in spatial dimension, respectively, and ${a}_{i}\equiv \partial_{i}\ln {N}$ with ${N}$ being a lapse function of the ADM line element. Here the $\alpha$, $\beta$ and $\lambda_{HL}$ are the dimensionless parameters of the effective HL theory in the IR limit. In addition, we used the middle Latin indices for labeling the spatial dimensions, i.e., $i,j,k = 1,2,3$. Here we start with the action in the Jordan frame with the non-minimal coupling, the $\xi\,\Phi^{2/d}\,\mathcal{R}$ coupling introduced in the action in Eq.(\ref{ADM-non-project-HL}), then the Jordan frame of the effective HL theory in IR limit is written by
\begin{eqnarray}
S &=& \int dt\,d^3x\,N\,\sqrt{h}\,\Big[ -\frac{1}{2}\,M_P^2\,\mathcal{R} - \frac{\xi}{2}\,\Phi^{2/d}\,\mathcal{R} + g^{\mu\nu}\Phi^{(2-2d)/d}\,\partial_\mu\,\Phi\,\partial_\nu\,\Phi - V^{IR}_{\rm eff}(\Phi)\Big]
\nonumber\\
&=& \int dt\,d^3x\,N\,\sqrt{h}\,\Big[ -\,\frac{1}{2}\,\Omega^{2}(\Phi)\,\mathcal{R} + g^{\mu\nu}\Phi^{(2-2d)/d}\,\partial_\mu\,\Phi\,\partial_\nu\,\Phi - V^{IR}_{\rm eff}(\Phi)\Big]\,,
\label{jordan-ADM-non-project-HL}
\end{eqnarray}
where
\begin{eqnarray}
M^{2}_{P}=(8\pi G)^{-1},\qquad\Omega^{2}\equiv \frac{M_P^2 + \xi\,\Phi^{2/d}}{M_P^2},
\end{eqnarray}
and 
\begin{eqnarray}
\mathcal{R} = (1 - \alpha)\,K_{ij}\,K^{ij} - (1 + \lambda_{HL})\,K^2 +\, {}^{(3)}R + \beta\,a_i\,a^i\,.
\end{eqnarray}
We use the conformal transformation technique from Refs.\cite{Kluson:2011qv} in order to transform the original Jordan-frame action to the Einstein frame. The IR effective action of the HL theory in the healthy extension \cite{Blas:2009qj} at $z\rightarrow 1$ limit in the Einstein frame without the cosmological constant takes the form:
\begin{eqnarray}
S_E = \int d^4x\sqrt{-\widetilde g}\,\Big[-\frac{1}{2}\,M_P^2\,\mathcal{\widetilde R} + \frac{1}{2}\widetilde g^{\mu\nu}\,\partial_\mu \chi\, \partial_\nu \chi - U(\chi)  \Big]\,,
\label{scalar-E-frame}
\end{eqnarray}
where the four dimensional Ricci-like scalar term ($\mathcal{\widetilde R}$) of HL gravity is written as
\begin{eqnarray}
\mathcal{\widetilde R} = (1-\alpha)\,\widetilde{K}_{ij}\widetilde{K}^{ij} + (1+ \lambda_{HL})\, \widetilde{K}^2 + \, {}^{(3)}\widetilde R + \beta\,\widetilde a_i\, \widetilde a^i\,.
\end{eqnarray}
Here the tildes represent the quantities in the Einstein frame. According to the parametrized post-Newtonian (ppN) constraints, see Refs.\cite{Gumrukcuoglu:2017ijh,Ramos:2018oku}, the coefficient $\beta$ accompanying $a_{i}a^{i}$ must be $\beta < 10^{-4}$ and a large value of this coefficient is also excluded by cosmological constraint. Nevertheless, the term with $\beta$ parameter will emerge when we perform perturbation or instability effects. In addition, Refs.\cite{Gumrukcuoglu:2017ijh,Ramos:2018oku} also constrained the bounds of the $\alpha$ parameter of the IR limit of the HL theory to $|\alpha|\lesssim 10^{-15}$. In this work, however, we focus on the study of background dynamics and therefore, the ``healthy extension" terms with parameters $\alpha,\,\beta$ can be neglected in our study. Having used all assumption above, one notes that the action in Eq.(\ref{scalar-E-frame}) is equivalent to the standard non-minimal Higgs inflation in the context of the ADM formalism. In order to obtain the Higg-like effective potential in Eq.(\ref{Higgs-like-pot}), in addition, the almost constant value of the $\varphi$ corresponds to the slow-roll approximation of the inflation and we will use this approximation in the latter. The kinetic term of the scalar field in the Einstein frame is related to that of the Jordan frame via \cite{Bezrukov:2011mv}
\begin{eqnarray}
\widetilde g^{\mu\nu}\partial_\mu \chi \partial_\nu \chi = \left( \frac{d\chi}{d\Phi}\right)^2 g^{\mu\nu}\partial_\mu \Phi \partial_\nu \Phi \,,
\end{eqnarray}
with
\begin{eqnarray}
\left( \frac{d\chi}{d\Phi}\right)^2 = 2\Omega^{-2}\left( 1 + \frac{2\xi^2}{d^2M_P^2}\Omega^{-2}\Phi^{\frac{2}{d}}\right)\Phi^{\frac{(2-2d)}{2d}} \,.
\end{eqnarray}
According to the standard composite NJL model, it was pointed out in Ref.\cite{Channuie:2016iyy} that the radial (real) part, $\varphi(x)$ of the composite fields is a dynamical field in the inflation phase. In this work, we will replace the scalar field $\Phi$ in Eq.(\ref{scalar-E-frame}) by the (re-scaled) composite scalar field $\varphi$ as
\begin{eqnarray}
\Phi \rightarrow \varphi \,.
\end{eqnarray}
This means that $d=1$ is implied and we have neglected a phase here. By employing the Higgs-like form of the composite effective potential near IR fixed point in Eq.(\ref{Higgs-like-pot}), the potential of the normalized scalar field in the action (\ref{scalar-E-frame}) is given by
\begin{eqnarray}
U(\chi) = \Omega^{-4}V_{\rm eff}^{IR}\big(\Phi(\chi)\big)\,,
\end{eqnarray}
and
\begin{eqnarray}
\Omega^2 = \frac{M_P^2 + \xi\varphi^2}{M_P^2} \,.
\end{eqnarray}
In the Einstein frame with a flat FLRW background, we obtain the HL gravity with composite (real) scalar field ($\varphi$) \cite{Kiritsis:2009sh,Calcagni:2009qw,Lu:2009em}. The relevant dynamics reads
\begin{eqnarray}
&& H^2 = \frac{1}{3 M_P^2 \kappa}\left( \frac{\kappa}{4}\dot\chi^2 + U\right)\, \label{H2},\\
&& H^2 + \frac23\dot H = -\frac{1}{3 M_P^2 \kappa}\left( \frac{\kappa}{4}\dot\chi^2 - U\right) \,,\label{Hdot}\\
&& \ddot\chi + 3 H \dot \chi + \frac{2}{\kappa}\,\frac{d U}{d\chi} = 0 \,,
\label{KG1}
\end{eqnarray}
where $\kappa = (3\lambda_{HL}-1)$\,. The slow roll parameters, $\epsilon,\eta,\zeta$, and the number of e-folding, $N$, have the same forms as the standard slow-roll paradigm, and they read
\begin{eqnarray}
\epsilon &=& \frac{M_P^2}{2} \left( \frac{1}{U}\frac{dU}{d \chi} \right)^2,\,
\qquad\quad
\eta = M_P^2 \left( \frac{1}{U}\frac{d^2U}{d \chi^2} \right), \\
\zeta &=& M_P^2 \left( \frac{1}{U^2}\frac{dU}{d \chi}\frac{d^3U}{d \chi^3} \right),\,
\qquad\quad
N = \frac{1}{M_P^2} \int _{\chi_{\rm end}} ^{\chi_N} d \chi  \frac{U}{dU/d\chi}\,.
\label{slow-roll}
\end{eqnarray}
Note that the inflaton field would be large at the inflationary scale. By using Eq.(\ref{H2})-Eq.(\ref{KG1}) in addition, we see below that the slow-roll parameters in our model are independent of the $\lambda_{HL}$. However, the residual parameter from HL theory, a.k.a. $\alpha$ introduced in Eq.(\ref{cut-off-ansatz}), appears in the slow-roll parameters instead. Using the following large-field approximation,
\begin{eqnarray}
\xi\varphi^2 \gg M_P^2\,,\quad\rightarrow \quad
\Omega^2 \approx \frac{\xi\varphi^2}{M_P^2}\,,
\end{eqnarray}
we can write the solution for $\varphi$ in terms of $\chi$ as,
\begin{eqnarray}
\varphi \simeq \frac{M_P}{\sqrt{\xi}}\,\exp\left( \frac{\chi}{\sqrt{6}\,M_P}\right)\,.
\end{eqnarray}
Then the inflaton potential in the Einstein frame takes the following form:
\allowdisplaybreaks
\begin{eqnarray}
U = \Omega^{-4}\,V_{\rm eff} \approx   \lambda\, \varphi^4 \left(1+\frac{\xi  \varphi^2}{M_P^2}\right)^{-2}.
\label{Einstein-eff-pot}
\end{eqnarray}
Having used the effective potential in Eq. (\ref{Einstein-eff-pot}), the slow-roll parameters are given by
\begin{eqnarray}
\epsilon \simeq \frac{4 M_P^4}{3 \xi ^2 \varphi^4},\,\qquad\eta \simeq-\frac{4 M_P^2}{3 \xi  \varphi^2} ,\,\qquad \zeta\simeq \frac{16 M_P^2}{9 \xi ^2 \varphi^4}.
\end{eqnarray}
The inflation terminates at $\epsilon = 1$\,, and the end of the inflation gives the solutions for the field $\varphi$ as,
\begin{eqnarray}
\varphi_{\rm end} \simeq \frac{\sqrt{2}M_P}{\sqrt[4]{3} \sqrt{\xi }} \sim 1.07 \frac{M_P}{ \sqrt{\xi }}.
\end{eqnarray}
We also can simply determine the e-folding number, $N$, to obtain
\begin{eqnarray}
N = \frac{1}{M_P^2}\int_{\varphi_{\rm end}}^{\varphi_N}\frac{U}{d U/d\varphi}\left( \frac{d\chi}{d\varphi}\right)^2 d\sigma
\simeq \frac{3 \xi  \varphi_N^2}{4 M_P^2}-\frac{\sqrt{3}}{2}.
\end{eqnarray}
Substituting $\varphi_{\rm end}$ in the e-folding number equation, one obtains, for the large field approximation:
\begin{eqnarray}
\varphi_{N} \simeq \frac{2 \sqrt{N}M_P}{ \sqrt{3\xi }}\,,
\end{eqnarray}
with $\varphi_{N}\simeq9M_{P}/\sqrt{\xi}$ for $N=60$. We are now ready to compare our predictions with experiments. We start by considering the constraints set by the observed amplitude of the density
perturbation, $A_{s}$. To generate the proper value of $A_{s}$, the potential must
satisfy at horizon crossing $\varphi=\varphi_{N}$:
\begin{eqnarray}
\left|U/\epsilon\right|_{\varphi_{N}} = (0.0269)^{4} M_{P}^{4}.\label{three}
\end{eqnarray}
This is used to impose a constraint on the self coupling, which must be unnaturally small in a minimally coupled quartic potential: $\lambda \sim 10^{-13}$ \cite{Guth:1982ec}. However, in the present case, Eq.(\ref{three}) yields the relations among $\xi,\,\lambda $ and $N$. We can solve for $\xi$ to obtain:
\begin{eqnarray}
\xi\simeq 1515.8\,N\sqrt{\lambda}.
\label{xi-N}
\end{eqnarray}
\begin{figure}[H]
\centering
	\includegraphics[width=.6\textwidth]{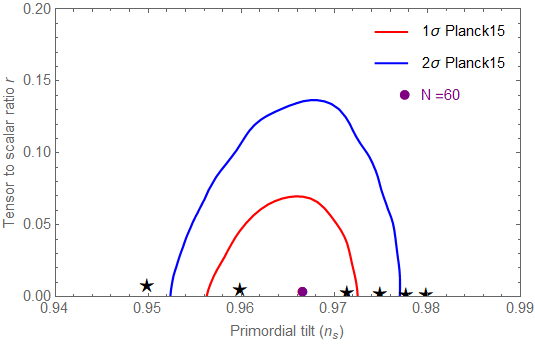}
	\caption{The theoretical predictions in the $(r-n_{s})$ plane for the present work  with Planck'15 results for TT, TE, EE, +lowP and assuming $\Lambda$CDM + r \cite{Ade:2015lrj}.}
	\label{rns}
\end{figure}
It is worth noting from Eq.(\ref{xi-N}) that the nonminimal coupling $\xi$ depends on both $N$ and $\lambda$. The later also depends on the HL parameter $\alpha$ defined in Eq.(\ref{couplings-higgs-pot}). For the up to two-loop contribution, at very high energy, a.k.a. an inflationary (grand unified energy) scale, the value of the coupling $\lambda$ approaches zero \cite{Degrassi:2012ry}. Here we assume for the one-loop level the value of $\lambda=0.001$, and compute the non-minimal coupling to approximately be $\xi\approx 9100$. We obtain the value very similar to that of the Higgs inflation \cite{Bezrukov:2007ep}. However, our effective theory for the composite inflaton cannot be used for arbitrary large value of the inflaton field. Here the effective theory is valid for $\sigma_{N} < 4\pi v$, implying for $N=60$ that
\begin{eqnarray}
v>\frac{9M_{P}}{4\pi\sqrt{\xi}}.
\label{vev}
\end{eqnarray}
With the given value of $\xi\sim 9100$ and Eq.(\ref{vev}), we find,
\begin{eqnarray}
v \sim 7.5 \times 10^{-3} M_{P} \approx 10^{16}~ {\rm GeV},
\label{vev-gut}
\end{eqnarray}
where we have employed the reduced Planck mass value. Interestingly, we also get
\begin{eqnarray}
\alpha \sim 100,
\label{alpha-100}
\end{eqnarray}
which allows to connect the two scales between IR and UV. From Eq.(\ref{couplings-higgs-pot}), by using $\lambda=0.001$, we obtain $g\sim 400$. This nicely corresponds to $v \sim 10^{-2} M_{P}$. Notice that the confining scale of the underlying theory is naturally of the order of the grand unified energy scale. The coincident value of $v\sim 10^{-2}M_P$ and $\alpha = 100$ is remarkably obtained in this work. According to the results in Eqs. (\ref{vev-gut}) and (\ref{alpha-100}), we argue that LI emerges firstly at the grand unified energy scale ($\sim 10^{16}$ GeV) in the framework of the HL theory. This result is also consistent to the conclusion of Refs. \cite{Kiritsis:2009sh,Calcagni:2009qw} that the universe flavors the LI in the inflationary phase in order to reproduce a correct power spectrum.

We can also test our predictions with the experimental results by using the relative strength of the tensor perturbation, i.e. the tensor-to-scalar ratio $r$, the spectral index of curvature perturbation $n_{s}$ and its running spectral index $n'_{s}$. In terms of the slow-roll
parameters, these observables are defined by
\begin{eqnarray}
r&=&16\,\epsilon_{\sigma_{N}},\,\quad n_{s}=1-6\,\epsilon_{\sigma_{N}}+2\eta_{\sigma_{N}},\,\\n'_{s}&=&-24\,\epsilon^{2}_{\sigma_{N}} + 16\epsilon_{\sigma_{N}}\eta_{\sigma_{N}}-2\zeta_{\sigma_{N}}.
\end{eqnarray}
To the lowest order in the slow-roll approximation, the inflationary predictions in terms of the
number of e-foldings in the Einstein frame parameters
for this model read: $n_{s}\simeq 1-2/N,\,n'_{s}\simeq 9/(2N^{4})-12/N^{3},\,r\simeq 12/N^{2}$. 
Specifically, we obtain for this model $n_{s}=0.96667,\,r=0.00333$ which perfectly lie inside the
$2\sigma$ region of the contours for $N=60$ e-foldings and $N_{c}=3$ as displayed in Fig.\ref{rns}. Concerning
the running spectral index $n'_{s}$, we obtain: $n'_{s}=-0.000055$ for $N=60$ e-foldings.
\begin{figure}[H]
\centering
	\includegraphics[width=.6\textwidth]{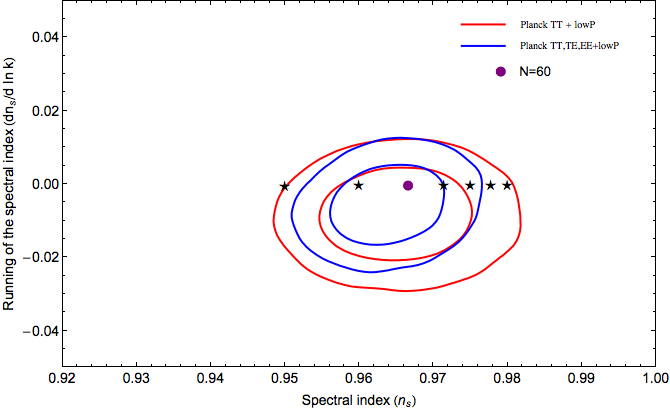}
	\caption{Marginalized joint $68\%$ and $95\%$ C.L. for $(n_{s}, n'_{s})$ using Planck TT+lowP and Planck TT,TE,EE+lowP \cite{Ade:2015lrj}.}
	\label{nsp}
\end{figure}
For this model, we notice that the running of the scalar spectral
index is nearly constant as a function of $n_{s}$ (see Fig.(\ref{nsp})).

\section{Conclusions}
In the present work, we have considered the so-called HL gravity in the context of inflation with our aim being dealing with the quantum theory of gravity. A basic assumption of the theory is an anisotropic scaling, $b$, of space and time in the UV regime, i.e, $\vec x\rightarrow b \vec x$ and $t\rightarrow b^{z}t$ with $z$ being the dynamical critical exponent. Clearly, it is realized that at very high energy the theory displayed a breaking of LI with $z=3$. However, at lower energy scale, the broken Lorentz symmetry can be recovered at $z=1$.

In the present work, we propose a cosmological scenario inherently based on the effective NJL model near the IR limit of the HL theory. In our scenario, the scalar channel of the NJL model plays a role of the composite inflaton, while the pseudo scalar is supposed to be a massless mode and decouples from the very high (inflationary) energy scale. We also discover the symmetry breaking of the inflaton potential near the $z=1$ limit, and assume the inflation starts at that IR regime in which the LI emerges. Here we introduced the effective cut-off, $M$, where this parameter appears slightly above the IR scale. We also demonstrate how the effective potential of the underlying theory can be obtained at $z=1$ limit, and show how to derive a Higgs-like potential. Here the symmetry of the effective potential will be broken near $z=1$ at certain value of the critical coupling; while at $z=3$ the symmetry of the effective potential will always be broken at arbitrary value of the coupling. We argue that this furnishes some hints about the emergence of LI and the broken phase of the symmetry with the given critical value. The dynamical origin of this point is worth further investigation. 

In order to demonstrate the emergence of LI at the IR fixed point in this scenario, we addressed this point by considering the inflationary phase of the universe as a probe of physics of the IR. This might provide a connection to the UV of the HL theory using the ansatz in Eq.(\ref{cut-off-ansatz}). In so doing, we follow the standard slow-roll scenario by computing the inflationary parameters and compare them with recent Planck 2015 data. Interestingly, we discovered that the predictions of the model are in good agreement with the Planck analysis. We notice, however, that the nonminimal coupling does implicitly depend on the parameter, $\alpha$, of the HL theory. By comparing the predictions with the data, we constrain the cosmological parameters in the HL theory both in the IR and UV limits, and discover $\alpha \sim 100$. Assuming a UV fixed point as Planck scale, our results 
%
%
suggested that the emergence of LI takes place at the grand unified energy scale. Our salient feature is that we used inflation to demonstrate that the grand unified energy scale is the IR fixed point of HL gravity, while the Planck scale is the UV fixed point of the theory. Note that these two scales can naturally be obtained in our framework. According to the results in this work, we found that the UV and IR regimes of the HL theory can be identified by using the inflationary cosmology constraints.

\acknowledgments

We thank Prof.\,Gregory W. Horndeski for his thorough readings and intuitive comments on our manuscript. We are also graceful to the anonymous referees for valuable comments and intuitive suggestions that helped us improve our manuscript. DS is supported by Thailand research fund (TRF) under contract No. TRG6180014. PC is financially supported by the Institute for the Promotion of Teaching Science and Technology (IPST) under the project of the \lq\lq Research Fund for DPST Graduate with First Placement\rq\rq\,, under Grant No. 033/2557. This research was partially supported by the New Strategic Research (P2P) project, Walailak University, Thailand.

\end{document}